\documentclass[pdflatex,sn-mathphys-num]{sn-jnl}% Math and Physical Sciences Numbered Reference Style 
\usepackage{multirow}%
\usepackage{amsmath,amssymb,amsfonts}%
\usepackage{amsthm}%
\usepackage{mathrsfs}%
\usepackage[title]{appendix}%
\usepackage{xcolor}%
\usepackage{textcomp}%
\usepackage{manyfoot}%
\usepackage{booktabs}%
\usepackage{algorithm}%
\usepackage{algorithmicx}%
\usepackage{algpseudocode}%
\usepackage{listings}%
\usepackage{soul}
\begin{document}
\title[Article Title]{A Higher-Order Poincar\'e Ellipsoid representation for elliptical vector beams}
%\definecolor{Dayver}{RGB}{0, 0, 255}
\author[1]{\fnm{Dayver} \sur{Daza-Salgado}}
\author[1]{\fnm{Edgar} \sur{Medina-Segura}}
%\equalcont{These authors contributed equally to this work.}
\author*[2]{\fnm{Valeria} \sur{Rodriguez-Fajardo}}
\email{vrodriguezf@unal.edu.co}
%\equalcont{These authors contributed equally to this work.}
\author*[3]{\fnm{Benjamin} \sur{Perez-Garcia}}
\email{b.pegar@tec.mx}
%\equalcont{These authors contributed equally to this work.}
\author*[1]{\fnm{Carmelo} \sur{Rosales-Guzmán}}\email{carmelorosalesg@cio.mx}
%\equalcont{These authors contributed equally to this work.}
\affil[1]{\orgdiv{}\orgname{Centro de Investigaciones en Óptica A.C.}, \orgaddress{\street{Loma del Bosque 115, Colonia Lomas del Campestre}, \city{León}, \postcode{37150}, \state{Guanajuato}, \country{México}}}
\affil[2]{\orgdiv{Departamento de Física}, \orgname{Universidad Nacional de Colombia Sede Bogotá}, \orgaddress{\street{Carrera 30 No. 45-03}, \city{Bogotá}, \postcode{111321}, \country{Colombia}}}
\affil[3]{\orgdiv{Photonics and Mathematical Optics Group}, \orgname{Tecnologico de Monterrey}, \orgaddress{ \city{Monterrey}, \postcode{64849}, \state{Nuevo León}, \country{México}}}

%%==================================%%
%% Sample for unstructured abstract %%
%%==================================%%

\abstract{The Higher-Order Poincaré Sphere (HOPS) provides a powerful geometrical tool for representing vector beams as points on the surface of a unitary sphere. Since a particular position on the surface represents any spatial mode regardless of its shape, this representation cannot be used to discern between the spatial modes geometries of vector modes. For instance, Laguerre- and Ince-Gauss vector beams are ambiguously represented using the same unitary sphere, even though their spatial profiles are circular and elliptical, respectively. As such, in this manuscript, we propose a generalisation of the HOPS that we call the Higher-Order Poincaré Ellipsoid (HOPE). Our approach allows an unambiguous representation of helical Ince-Gauss vector modes of ellipticity $\varepsilon$ onto the surface of an ellipsoid of eccentricity $\bf e$, providing a unique way to visualise elliptically-shaped vector modes. We provide a transformation that links the ellipticity $\varepsilon$ of helical Ince-Gauss vector modes to the eccentricity $\bf e$ of an ellipsoid, such that the HOPS is recovered for $\varepsilon=0$. Since this representation preserves the Stokes parameters formalism, the transition from the HOPS to the HOPE is straightforward, thus making its implementation appealing for the structured light community. We anticipate the concepts outlined here will pave the path toward the representation of structured light beams' properties using other geometrical objects.}

\keywords{Structured light, vector beams, Ince-Gauss beams, Higher-order Poincar\'e sphere, inhomogeneous polarisation, Stokes parameters.}
\maketitle

%%%%%%%%%%%%%%%%%%%%%%%%%%%%%%%%%%%%%%%%%%%%%%%%%%
\section{Introduction}\label{sec1}
%%%%%%%%%%%%%%%%%%%%%%%%%%%%%%%%%%%%%%%%%%%%%%%%%%
Structured light beams, in which one or more degrees of freedom are engineered to generate light beams with novel properties, represent one of the most general states of light \cite{Roadmap,Forbes2021StructuredLight}. Typical examples of structured beams are solutions to the wave equation in its exact or paraxial form such as Hermite-, Laguerre- and Ince-Gauss (IG), which correspond to solutions in cartesian, circular cylindrical, and elliptical cylindrical coordinates, respectively \cite{siegman,bandres2004-Ince,SPIEbook}. The IG family is characterised by an ellipticity parameter $\varepsilon\in [0,\infty)$ that provides a smooth transition from the Laguerre- with $\varepsilon=0$ to the Hermite-Gauss modes as $\varepsilon\to\infty$ \cite{wang2022exploring,Bandres2004}, and has found applications in quantum mechanics \cite{gonzalez2023nonlocality}, optical communications \cite{zhu2020entanglement}, optical trapping \cite{otte2020optical}, among others \cite{wang2022exploring}. It is also possible to tailor both the spatial and polarisation degrees of freedom (DoF), thereby producing vector modes. These are characterised by a non-homogeneous and exotic polarisation distribution and are generated as a non-separable superposition of orthogonal polarisations and orthogonal spatial modes \cite{Galvez2012}. Vector modes are pioneering a wide variety of applications across different fields \cite{Rosales2018Review,rosales2024perspective,Ndagano2018,Hu2024,Hu2019,yang2021,BergJohansen2015}, and therefore they have become an important field of research. Importantly, while the polarisation basis has only two values, the spatial mode basis has infinite elements. They are often regarded as the classical analogous of quantum-entangled states since both are expressed mathematically as a non-separable product. This mathematical similarity has coined the very controversial name ``classically-entangled'' modes \cite{Karimi2015,konrad2019quantum,Paneru2020}. Contrary to quantum entanglement, classical entanglement is of local nature, where every photon in a classical beam of light is said to be entangled in two DoFs. Similarly to entangled photons, in vector modes it is not possible to measure the polarisation DoF without influencing the spatial one and vice versa. This property represents an advantage over some conventional scalar optical fields and, as such, it has been applied in different areas of science, technology, engineering, and industry, where traditional optical fields have encountered physical limitations \cite{Ndagano2017,Shen2022,Toppel2014,forbes2019classically}. It is also worth mentioning that vector beams are resilient to perturbations in unitary complex channels where one of the two DoFs is unaffected, such as atmospheric turbulence \cite{nape2022}.

Since the number of spatial modes is unbounded, it is possible to generate an infinite number of vector beams 
from all possible combinations of orthogonal spatial modes and orthogonal polarisation states \cite{dudley2013generating,Zhaobo2022,HuXiaobo2021,Hu2022abruptly,Mathieu_VB_2021,Medina-Segura:2023,rodriguez2024experimental}. In addition, for a particular combination of spatial modes and polarisation, there is an infinite set of vector states obtained via two characteristic parameters that can be changed continuously: the intermodal phase and the weighting coefficient. The former is a phase factor that rotates the polarisation state along the transverse plane. The latter is a weighting factor, used to tune the coupling between both DoFs or the ``vectorness" of the beam and allows a transition of the same from purely scalar to purely vectorial. The infinite combinations can be conveniently represented on the surface of a unitary-radius sphere known as the Higher-Order Poincar\'e Sphere (HOPS) \cite{milione2011, holleczek2011classical}. In this representation, scalar modes with right- and left-handed circular polarisation are represented on the north and south poles, respectively, the pure vector ones on the equator and the rest on the remaining points of the sphere \cite{Ndagano2016,Zhaobo2020,McLaren2015}. This representation is used for any vector mode, regardless of its spatial shape. For instance, helical-Ince Gauss (hIG) vector modes with the same transverse parameters and different $\varepsilon$ values have distinct shapes, yet they are ambiguously located on the same point on the HOPS \cite{Yao-Li2020}. It is worth mentioning that there are other geometric representations for vector modes, such as the generalised Poincar\'e sphere, which links the radius of the Poincar\'e sphere to a continuous change of polarisation ellipticity \cite{ren2015generalized} or the five-dimensional Poincar\'e sphere system, which generalises the previous one \cite{zhao2022five}. Crucially, none of these representations account for the spatial shape of the modes as they only consider information about polarisation.

In this manuscript, we propose a geometric representation of vector modes, which we apply to hIG vector beams but can be extended to other vector beams. In this representation, we link the ellipticity parameter $\varepsilon$ to the eccentricity $\bf e$ of an ellipsoid by expressing the first in terms of the latter, such that all IG vector modes of ellipticity $\varepsilon$ are represented onto the surface of the ellipsoid. Essentially, we are generalising the HOPS into what we have termed the High-order Poincaré Ellipsoid (HOPE). It is worth emphasising that the relation between the Stokes parameters and the Cartesian coordinates is preserved. Additionally, for $\varepsilon=0$ the HOPE transforms into the HOPS, as desired. The HOPE we propose here provides an intuitive geometric description of the transformation of helical Ince-Gauss vector modes in both DoFs, polarisation, and spatial shape. In the former, from circular to linear polarisation, and in the latter from circular to rectangular shape. We thus anticipate that the HOPE will become a useful tool for the structured light community. 

%%%%%%%%%%%%%%%%%%%%%%%%%%%%%%%%%%%%%%%%%%%%%%%%%%
\section{Mathematical Framework}
%%%%%%%%%%%%%%%%%%%%%%%%%%%%%%%%%%%%%%%%%%%%%%%%%%

%%%%%%%%%%%%%%%%%%%%%%%%%%%%%%%%%%%%%%%%%%%%%%%%%%
\subsection{Scalar helical Ince-Gauss modes}
%%%%%%%%%%%%%%%%%%%%%%%%%%%%%%%%%%%%%%%%%%%%%%%%%%
The Ince-Gauss modes constitute an infinite set of orthogonal solutions to the paraxial wave equation (PWE) in elliptical-cylindrical coordinates ${\bf r}=(\xi,\eta,z)$, where $\xi\in[0,\infty)$ and $\eta\in[0,2\pi)$ are the radial and angular elliptical coordinates, respectively, and $z$ is the propagation coordinate. IG modes are described mathematically as
	\begin{align}
		\text{IG}_{p,m;\varepsilon}^{e}({\bf r})&=\frac{\mathcal{C}\omega_0}{\omega(z)}C_p^m(i\xi;\varepsilon)C_p^m(\eta;\varepsilon)\exp{\left[\frac{-\rho^2}{\omega^2(z)} \right]}\exp\left[i\left(kz+\frac{k\rho^2}{2R(z)}-\zeta(z)\right) \right], \label{2}\\
        \text{IG}_{p,m;\varepsilon}^{o}({\bf r})&=\frac{\mathcal{S}\omega_0}{\omega(z)}S_p^m(i\xi;\varepsilon)S_p^m(\eta;\varepsilon)\exp{\left[\frac{-\rho^2}{\omega^2(z)} \right]}\exp{\left[i\left(kz+\frac{k\rho^2}{2R(z)}-\zeta(z) \right)\right],} \label{3}
	\end{align}
 where the superscripts $e$ and $o$ denote the even and odd solutions, respectively, $\mathcal{C}$, $\mathcal{S}$ are constants of normalisation \cite{bandres2004-Ince}, and $\rho=\sqrt{x^2+y^2}$ is the magnitude of position vector in the transverse plane. Here, $C_p^m$ and $S_p^m$ are the even and odd Ince polynomials, respectively, with $\varepsilon\in[0,\infty)$ the ellipticity parameter, and $0\leq m\leq p$ for even functions and $1\leq m\leq p$ for odd functions, such that $p$ and $m$ must have the same parity, i.e. $(-1)^{p-m}=1$. The index $m$ corresponds to the number of hyperbolic nodal lines, while the index $p$ to the number of elliptic nodal lines. In addition, $R(z)=z+z_R^2/z$ is the radius of curvature of the wavefront, where $z_R=\pi\omega_0^2/\lambda$ is the Rayleigh distance, $\omega(z)=\omega_0\sqrt{1 + \left(z/z_R\right)^2}$ is the beam size with a minimum waist $\omega_0$ at the plane $z=0$, $k=2\pi/\lambda$ is the wave number, and $\zeta(z)=(p+1)\arctan(z/z_R)$ is the Gouy phase.
 
\begin{figure}[h]
	\centering
	\includegraphics[width=1\textwidth]{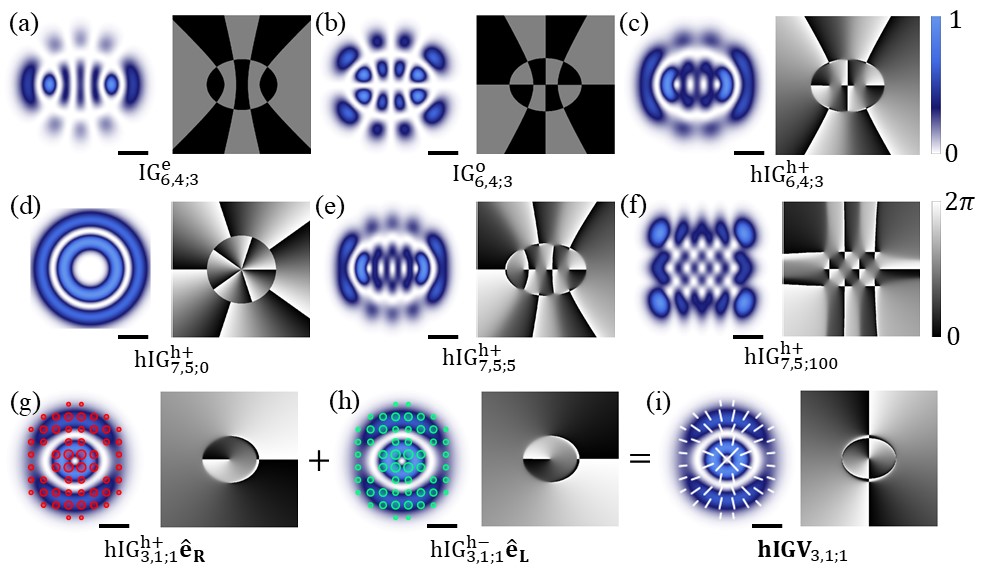}
	\caption{Theoretical intensity (left) and phase (right) distribution of even (a), odd (b) and helical (c) IG beams of parameters $m=4$, $p=6$ and ellipticity $\varepsilon=3$. Similarly, for hIG modes with parameters $p=7$, $m=5$ and ellipticities (d) $\varepsilon=0$, (e) $\varepsilon=5$, and (f) $\varepsilon=100$. The last row shows the schematic representation of the non-separable superposition to generate hIG vector modes. Left panels show the intensity distributions overlapped by the transverse polarisation distribution. The red and green circles represent right- and left-handed circular polarisation, respectively, and the white lines represent linear polarisation. The right panels show the phase distributions of the scalar modes ((g) and (h)), and the phase of the Stokes field $S = S_1 + i S_2$ (i). For this example we used the parameters $p=3$, $ m=1$, $\varepsilon=1$, $\theta=\pi/4$, and $\delta=0$. The scale of the beams is 1 mm and is represented by the black right lower bar.} 
\label{fig:figure1}
\end{figure}

A coherent superposition of the even and odd IG modes gives rise to the  helical Ince-Gauss (hIG) modes defined as,
\begin{equation}\label{helicalIG}
    \text{hIG}_{p,m;\varepsilon}^{h\pm}({\bf r})=\text{IG}_{p,m;\varepsilon}^{e}({\bf r})\pm i\, \text{IG}_{p,m;\varepsilon}^{o}({\bf r})\,,    
\end{equation}
where the sign $h+$ or $h-$ stands for modes with positive or negative helicity, respectively. Notably, hIG modes are only defined for $m>0$ because the odd solution $\text{IG}_{p,m;\varepsilon}^o$ is not defined for $m=0$. By way of example, the top row of Fig.~\ref{fig:figure1} shows the theoretical intensity (left) and phase distribution (right) of an even (Fig.~\ref{fig:figure1}(a)), odd (Fig.~\ref{fig:figure1}(b)) and helical (Fig.~\ref{fig:figure1}(c)) IG modes for the specific parameters $m=4$, $p=6$ and ellipticity  $\varepsilon=3$, at the plane $z=0$. Furthermore, the ellipticity parameter $\varepsilon$ allows the transition of IG modes from Laguerre-Gauss ($\varepsilon$=0), to Hermite-Gauss modes ($\varepsilon\to\infty$) \cite{bandres2004-Ince}, as illustrated in the middle row of Fig.~\ref{fig:figure1}. Here, the theoretical intensity (left) and phase (right) profiles of IG modes with parameters $p=7$ and $m=5$ for ellipticity values, $\varepsilon=0$ (Fig.~\ref{fig:figure1}(d)), $\varepsilon=5$ (Fig.~\ref{fig:figure1}(e)) and $\varepsilon=100$ (Fig.~\ref{fig:figure1}(f)) are shown. Note that for $\varepsilon=0$ the hIG modes acquire the characteristic spatial structure of the LG modes.

%%%%%%%%%%%%%%%%%%%%%%%%%%%%%%%%%%%%%%%%%%%%%%%%%%
\subsection{ Vectorial helical Ince-Gauss modes}
%%%%%%%%%%%%%%%%%%%%%%%%%%%%%%%%%%%%%%%%%%%%%%%%%%
In general, vector beams are regarded as a non-separable weighted superposition of two or more degrees of freedom \cite{Shen2022}. In particular, hIG vector modes, classically entangled in their spatial and polarisation degrees of freedom, can be expressed mathematically as \cite{SPIEbook},
\begin{equation}\label{hIGV}
    {\bf hIGV}_{p,m;\varepsilon}({\bf r}) = \cos\theta \, \text{hIG}_{p,m;\varepsilon}^{h+}({\bf r})\,
    \hat{{\bf e}}_{R} + \sin\theta \, {\text e}^{i\delta} \text{hIG}_{p,m;\varepsilon}^{h-}({\bf r})\,\hat{{\bf e}}_{L},
\end{equation}
%\begin{align}
	%{\bf E}({\bf r})= E_R({\bf r})\hat{{\bf e}}_R \cos\theta + {\text e}^{\imath\delta}E_L({\bf r})\hat{{\bf e}}_L \sin\theta,
	%\label{vectormode}
%\end{align}
where $\hat{{\bf e}}_R$ and $\hat{{\bf e}}_L$ are the unitary vectors representing right and left circular polarisation, respectively, with spatial distributions given by the scalar modes $\text{hIG}_{p,m;\varepsilon}^{h+}$ and $\text{hIG}_{p,m;\varepsilon}^{h-}$, respectively. The parameter $\delta$ $\in[0,\pi]$ is the intermodal phase that introduces a phase delay between both scalar modes and rotates the transverse polarisation distribution. The parameter $\theta\in[0,\pi/2]$ is a weighting factor that tunes the vectorness of the mode, from purely scalar ($\theta=0$ and $\theta=\pi/2$) to purely vectorial ($\theta=\pi/4$). The bottom row in Fig.~\ref{fig:figure1} schematically illustrates the superposition process given by Eq. \eqref{hIGV}. The polarisation distribution overlapped with the intensity profile (right) and the phase (left) of the scalar modes $\text{hIG}_{p,m;\varepsilon}^{h+}$ and $\text{hIG}_{p,m;\varepsilon}^{h-}$ are shown in Figs. \ref{fig:figure1}(g) and \ref{fig:figure1}(h), respectively. Similarly, the right panels of Fig.~\ref{fig:figure1}(i) shows the phase of the Stokes field $S = S_1 + i S_2$, whereas the left panel shows the intensity distributions overlapped by the inhomogeneous transverse polarisation distribution of the vector mode ${\bf hIGV}_{p,m;\varepsilon}({\bf r})$. Here, red and green circles represent right- and left-handed circular polarisation, respectively, whereas the white lines represent linear polarisation. For this examples we used the parameters $p=3$, $ m=1$, $\varepsilon=1$, $\theta=\pi/4$ and $\delta=0$.
 
 %\begin{figure}[ht]
%	\centering	\includegraphics[width=.85\textwidth]{figure 2.jpg}
%	\caption{Schematic representation of the non-separable superposition to generate hIG vector modes. The back panels show the phase distribution of the scalar modes ((a) and (b)), and the phase of the Stokes field $S = S_1 + i S_2$ (c). The front panels show the intensity distributions overlapped by the transverse polarisation distribution. The red and green circles represent right- and left-handed circular polarisation, respectively, and the white lines represent linear polarisation. For this example we used the parameters $p=3$, $ m=1$, $\varepsilon=1$, $\theta=\pi/4$, and $\delta=0$}
    %\label{fig:F3}
%\end{figure}

%%%%%%%%%%%%%%%%%%%%%%%%%%%%%%%%%%%%%%%%%%%%%%%%%%
\subsection{The Higher-Order Poincaré Ellipsoid}
\begin{figure}[ht]
	\centering	\includegraphics[width=.7\textwidth]{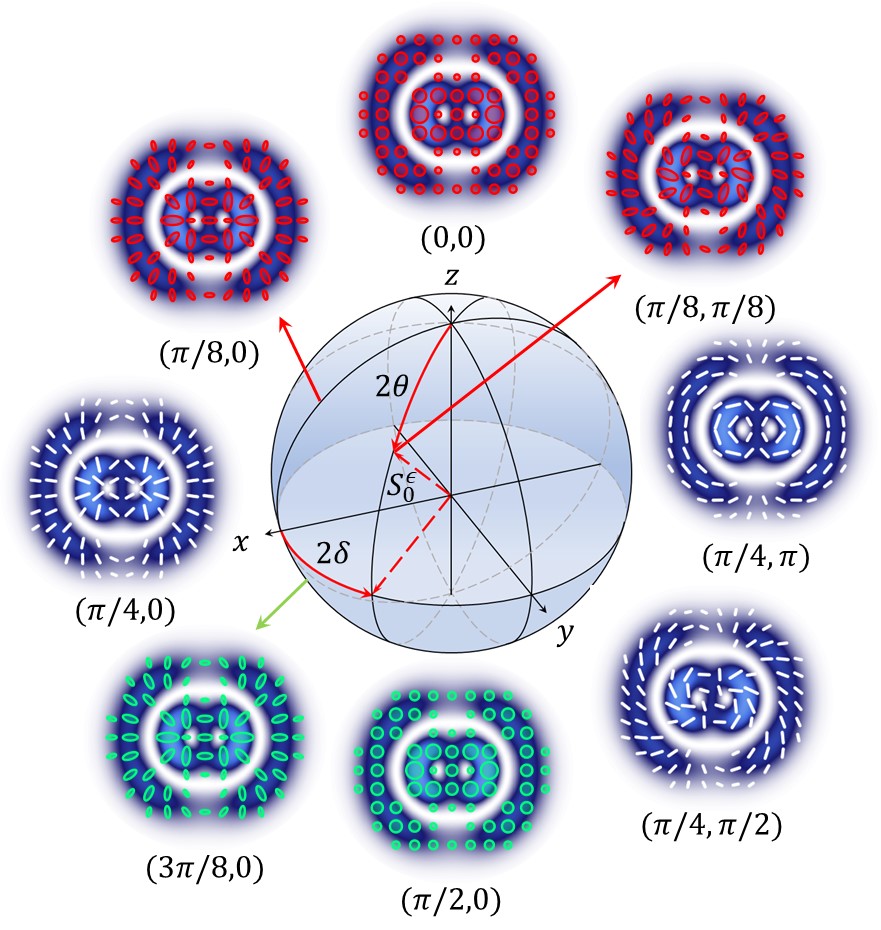}
	\caption{Traditional representation of hIG vector modes with parameters $p=4$, $m=2$ and ellipticity $\varepsilon=1$ on the surface of a HOPS. Here, scalar modes are represented on the poles, maximally non-separable modes are distributed along the equator and the rest, which carry elliptical polarisation, on the rest of the surface. The intensity and polarisation distributions of exemplary vector modes and their coordinates $(2\theta,2\delta)$ are also shown. Here, right- and left-handed elliptical polarisation states are represented by red and green ellipses respectively, whereas linear with white segments.}
\label{fig:HOPS}
\end{figure}

%%%%%%%%%%%%%%%%%%%%%%%%%%%%%%%%%%%%%%%%%%%%%%%%%%
As mentioned earlier, the HOPS \cite{milione2011, holleczek2011classical} is used to geometrically represent vector beams onto the surface of a unitary sphere. Examples of these are Laguerre- \cite{yi2015hybrid, marco2022extending, Hu2022Generation}, Ince- \cite{Yao-Li2020}, Mathieu- and Parabolic-Gaussian vector modes, among others \cite{perez2022highly,Hu2022Generation}. In this representation, the weighting coefficient -determined by $\theta$- and intermodal phase $\delta$ of the vector mode are associated with points $(2\theta,2\delta)$ on the surface of the HOPS. As an example, Fig.~\ref{fig:HOPS} shows the representation of a hIG vector mode of ellipticity $\varepsilon=1$ and parameters $p=4$, $m=2$. Notice that even though the shape of the represented mode is elliptical, its representation in the HOPS does not reflect in any way the shape of the mode. As such, in what follows we will introduce a generalisation -the HOPE-, which takes into account the shape of the mode.

A natural generalisation of a sphere is an ellipsoid with the canonical form,
\begin{equation}
    \frac{x^2}{a^2}+\frac{y^2}{b^2}+\frac{z^2}{c^2}=1.
    \label{eq:ellipse}
\end{equation}

%\begin{equation}
%    \frac{S_1^{\varepsilon^2}}{a^2}+\frac{S_2^{\varepsilon^2}}{b^2}+\frac{S_3^{\varepsilon^2}}{c^2}=S_0^{\varepsilon^2},
%\end{equation}
For convenience, we use an oblate spheroid, for which $a=b=1$ and define the parameter $c$ in terms of the eccentricity ${\bf e}$ of the spheroid as
\begin{equation}
\label{eq:eccentricity}
c=\sqrt{1-{\bf e}^2},
\end{equation}
{where, we relate the eccentricity ${\bf e}$ of the oblate spheroid and the ellipticity of the IG modes through the expression
\begin{equation}
    {\bf e}(\varepsilon):=\frac{\arctan \varepsilon}{\pi}.
    \label{eq:epsilo2ecc}
\end{equation}
In this way, for $\varepsilon=0$, ${\bf e}=0$ and the oblate ellipsoid becomes a sphere, thus recovering the conventional HOPS, as desired since for this value of $\varepsilon$ the IG become the LG vector beams. For $\varepsilon\rightarrow\infty$, the eccentricity ${\bf e}$ acquires the minimum value ${\bf e}=1/2$ and the surface transforms into a flattened ellipsoid. It is worth mentioning that Eq. \ref{eq:epsilo2ecc} was chosen in this way to highlight the asymptotic behaviour of IG Gauss modes as $\varepsilon\rightarrow\infty$. More precisely, even though theoretically HG modes are obtained when $\varepsilon\to\infty$, experimentally with $\varepsilon=100$ (${\bf e}=0.497$) we obtain a good approximation to HG modes. In this way, ${\bf e}\in[0,1/2)$ and therefore, $c\in[1, \sqrt{3}/2)$.} Hence, we can now represent hIG vector beams of ellipticity $\varepsilon$ onto the surface of oblate spheroids of eccentricity {\bf e}  with coordinates $(2\theta, 2\delta)$. We term this new representation the Higher-Order Poincar\'e Ellipsoid (HOPE).

%with ${\bf e}\in[0,0.5]$ to restrict $c$ to the values $[\sqrt{3/2}, 1]$. Even though the value ${\bf e}=0.5$ seems arbitrary, it was chosen so that the oblate spheroid does not become completely flat, when $c=0$, and also because . We now link the ellipticity $\varepsilon\in[0,\infty)$ of the IG modes to the eccentricity ${\bf e}$ of the spheroid through the relation, 

\begin{figure}[tb]
	\centering
\includegraphics[width=1\textwidth]{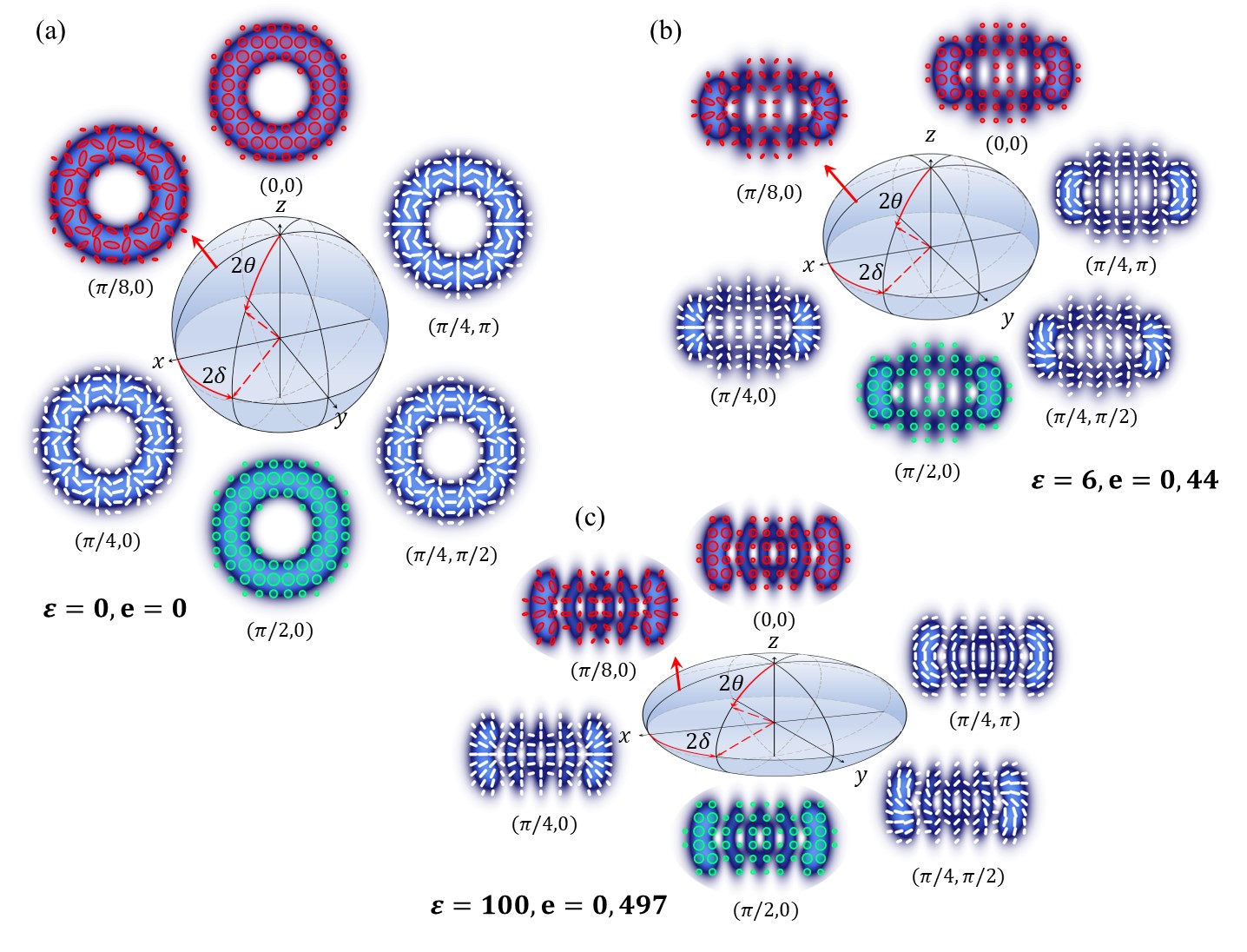}
	\caption{Geometric representation of hIG vector modes with parameters $m=5$ and $p=5$ on different HOPEs. Vector modes with ellipticity (a) $\varepsilon=0$ (b) $\varepsilon=6$ and (c) $\varepsilon=100$ are represented on the surface of a oblate spheroid of eccentricity ${\bf e}=0$, ${\bf e}=0.44$ and ${\bf e}=0.497$, respectively, the first corresponding the well-known HOPS.}
 \label{fig:trans}
\end{figure}

By way of example, Fig.~\ref{fig:trans} shows the representation of three hIG vector modes with parameters $p=5$ and $m=5$. First, in Fig.~\ref{fig:trans}(a) we show the case $\varepsilon=0$ (corresponding to a sphere, i.e. ${\bf e}=0$), which is precisely the well-known HOPS. The case $\varepsilon=6$ is illustrated in  Fig.~\ref{fig:trans}(b) (corresponding to an oblate spheroid of eccentricity ${\bf e}=0.44$), and Fig.~\ref{fig:trans}(c) shows the case $\varepsilon=100$ (corresponding to an eccentricity ${\bf e}=0.497$). Note also that similarly to the HOPS, scalar modes with right- and left-handed circular polarisation are located on the north and south poles, respectively, pure vector modes with a linear polarisation distribution are localized along the equator, and the rest of the modes are located on the remaining surface of the oblate spheroid.

It is also worth emphasizing that the Cartesian coordinates can be related to the Stokes parameters through the relations,
\begin{equation}\label{Eq:NewStokes}
    x=\frac{S_1^{\varepsilon}}{S_0^{\varepsilon}},\qquad
    y=\frac{S_2^{\varepsilon}}{S_0^{\varepsilon}},\qquad
    z=c\frac{S_3^{\varepsilon}}{S_0^{\varepsilon}}\,,
\end{equation}
such that the higher-order Stokes parameters can be written in terms of the amplitudes of the constituting scalar beams and the coordinates of the ellipsoid as
\begin{equation}
\begin{split}
    S_0^{p,m,\varepsilon}&=\cos^2{\theta}\iint_{-\infty}^\infty|\text{hIG}_{p,m;\epsilon}^{h+}({\bf r})|^2 dA + \sin^2{\theta}\iint_{-\infty}^\infty|\text{hIG}_{p,m;\epsilon}^{h-}({\bf r})|^2 dA = 1,\\
    S_1^{p,m,\varepsilon}&=2\text{Re}\left\{ \cos{\theta}\sin{\theta}e^{i\delta}\iint_{-\infty}^\infty|\text{hIG}_{p,m;\varepsilon}^{h+}({\bf r})|^2 dA \iint_{-\infty}^\infty|\text{hIG}_{p,m;\varepsilon}^{h-}({\bf r})|^2 dA\right\} = \sin{2\theta}\cos{\delta},\\
    S_2^{p,m,\varepsilon}&=2\text{Im}\left\{ \cos{\theta}\sin{\theta}e^{i\delta}\iint_{-\infty}^\infty|\text{hIG}_{p,m;\varepsilon}^{h+}({\bf r})|^2 dA \iint_{-\infty}^\infty|\text{hIG}_{p,m;\varepsilon}^{h-}({\bf r})|^2 dA\right\} = \sin{2\theta}\sin{\delta},\\
    S_3^{p,m,\varepsilon}&=\cos^2\theta\iint_{-\infty}^\infty|\text{hIG}_{p,m;\epsilon}^{h+}({\bf r})|^2 dA - \sin^2\theta\iint_{-\infty}^\infty|\text{hIG}_{p,m;\epsilon}^{h-}({\bf r})|^2 dA = \cos{2\theta},
    \label{Eq:Stokes2}
\end{split}
\end{equation}
%\begin{align}
%\begin{split}
    %S_0^{\varepsilon}&= |\text{hIG}_{p,m;\varepsilon}^{h+}({\bf r})|^2 +|\text{hIG}_{p,m;\varepsilon}^{h-}({\bf r})|^2 \\
%    S_0^{\varepsilon}&=\iint_{-\infty}^\infty|\text{hIG}_{p,m;\epsilon}^{h+}({\bf r})|^2 dA+ \iint_{-\infty}^\infty|\text{hIG}_{p,m;\epsilon}^{h-}({\bf r})|^2 dA,\\
%    S_1^{\varepsilon}&=S_0^{\varepsilon}\cos(2\theta)\cos(2\delta),\\
%    S_2^{\varepsilon}&=S_0^{\varepsilon}\cos(2\theta)\sin(2\delta),\\
%    S_3^{\varepsilon}&=S_0^{\varepsilon}\sin(2\theta)\,,
%\end{split}
%\end{align}
%or alternatively,
%\begin{equation}
%\begin{split}
%    S_0^{\varepsilon}&=\iint_{-\infty}^\infty|\text{hIG}_{p,m;\epsilon}^{h+}({\bf r})|^2 dA+ \iint_{-\infty}^\infty|\text{hIG}_{p,m;\epsilon}^{h-}({\bf r})|^2 dA,\\
%    S_1^{\varepsilon}&=2\sqrt{\iint_{-\infty}^\infty|\text{hIG}_{p,m;\varepsilon}^{h+}({\bf r})|^2 dA\iint_{-\infty}^\infty|\text{hIG}_{p,m;\varepsilon}^{h-}({\bf r})|^2 dA}\;\;\cos(\delta),\\
%    S_2^{\varepsilon}&=2\sqrt{\iint_{-\infty}^\infty|\text{hIG}_{p,m;\varepsilon}^{h+}({\bf r})|^2 dA\iint_{-\infty}^\infty|\text{hIG}_{p,m;\varepsilon}^{h-}({\bf r})|^2 dA}\;\;\sin(\delta),\\
%    S_3^{\varepsilon}&=\iint_{-\infty}^\infty|\text{hIG}_{p,m;\epsilon}^{h+}({\bf r})|^2 dA- \iint_{-\infty}^\infty|\text{hIG}_{p,m;\epsilon}^{h-}({\bf r})|^2 dA,
%    \label{Eq:Stokes2}
%\end{split}
%\end{equation}
which after a straightforward substitution of Eq. \ref{Eq:Stokes2} into Eq. \ref{Eq:NewStokes} and Eq. \ref{eq:ellipse} for $a=b=1$, results in the well-known relation  $(S_0^{\varepsilon})^2 = (S_1^{\varepsilon})^2 + (S_2^{\varepsilon})^2 + (S_3^{\varepsilon})^2$, for completely polarized optical fields.

%One of the key aspects of this geometric representation is that the eccentricity of the oblate spheroid, determined by the parameter $\bf e$, is directly linked to the ellipticity of the vector modes.

%%%%%%%%%%%%%%%%%%%%%%%%%%%%%%%%%%%%%%%%%%%%%%%%%%
\section{Experimental generation of helical Ince-Gauss modes}
\begin{figure*}[ht]
	\centering
\includegraphics[width=1\textwidth]{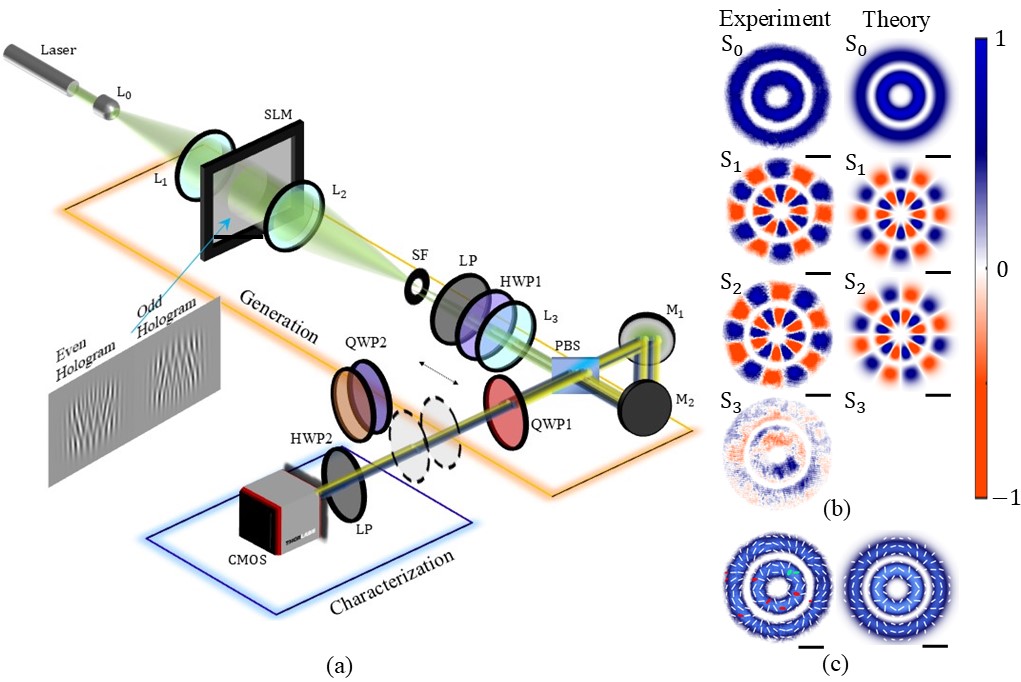}
	\caption{Experimental generation of vector modes. (a) Schematic of the experimental setup to generate and characterise vector modes. Here, L$_{i}$, $i=0,1,2,3$: lens; SLM: spatial light modulator; SF: spatial filter; LP: linear polariser; HWP, QWP: half and quarter wave-plate, respectively; PBS: polarising beam splitter; M$_i$, $i=1,2$: mirror; CMOS: Complementary Metal Oxide Semiconductor. The polarisation distribution of the vector modes was reconstructed through Stokes polarimetry. (b) Experimental and theoretical Stokes parameters of a hIG vector mode with $\varepsilon=0$, $m=3$ and $p=5$;  (c) the experimental (left) and theoretical (right) results for the reconstructed vector mode. The scale bar on the bottom of each figure is 1 mm.}
 \label{fig:exp}
\end{figure*}

%%%%%%%%%%%%%%%%%%%%%%%%%%%%%%%%%%%%%%%%%%%%%%%%%%
The experimental generation of hIG vector modes was implemented using the optical setup depicted in Fig.~\ref{fig:exp}(a). First, a linearly-polarised laser beam ($\lambda=532$ nm) is expanded by a microscope objective (L$_0$) and collimated by the lens L$_1$ of focal length $f_1=250$ mm, to obtain an approximate flat wavefront at the SLM plane. Afterward, the beam is sent through the screen of a transmission SLM (HOLOEYE LC 2012 with a resolution of 1024$\times$768 pixels, and a pixel pitch of 36 $\mu$m) where two independent computer-generated holograms are displayed side by side. Each hologram, computed following a complex amplitude modulation approach \cite{arrizon2007,SPIEbook}, contains the amplitude and phase information of orthogonal hIG modes. After the SLM, the first diffraction order of each mode is filtered with the help of a spatial filter (SF) located at the Fourier plane of the telescope formed by lens L$_2$ ($f_2=150$ mm) and L$_3$ ($f_3=175$ mm). The resulting beams are then transmitted through a half wave-plate oriented at 22.5$^\circ$ to rotate the polarisation of the beam from horizontal to diagonal. Both beams then enter a Sagnac interferometer through a polarising beam splitter (PBS) that transmits two horizontally polarised beams and reflects two vertically polarised beams. After a round trip, four beams exit the interferometer from the adjacent port. The two orthogonal spatial modes with orthogonal polarisation states are then superimposed to generate a vector beam, which is transformed to the circular polarisation basis using a quarter wave-plate at 45$^\circ$ \cite{Perez-Garcia2017}. The polarisation distribution of the modes was reconstructed through Stokes polarimetry with the help of a linear polariser, wave plates, and a CMOS digital camera, as explained below.

To reconstruct the polarisation distribution of the vector modes we used Stokes polarimetry, which is based on intensity measurements according to \cite{goldstein2017polarized},
\begin{align}
\begin{split}
    S_0=I_R+I_L,  &\qquad    S_1=2I_H-S_0,\\
    S_2=2I_D-S_0, &\qquad    S_3=2I_R-S_0,
\end{split}
\end{align}
where $I_H$, $I_D$, $I_R$, and $I_L$ are the horizontal, diagonal, right- and left-handed polarisation components, respectively, which can be measured experimentally following the technique described in \cite{Yao-Li2020}. By way of example, the left column of Fig.~\ref{fig:exp}(b) shows the Stokes parameters of an experimental hIG vector mode with parameters $m=3$, $p=5$, and $\varepsilon=0$. For comparison, the right column of Fig.~\ref{fig:exp}(b) shows its numerically simulated counterpart. The corresponding intensity profile overlapped by the transverse polarisation is shown in Fig.~\ref{fig:exp}(c) experiment on the left and numerical simulation on the right.
%\begin{figure}[ht]
%	\centering
%	\includegraphics[width=.8\textwidth]{figure 6.jpg}
%	\caption{Polarisation reconstruction of a {\bf hIGV} mode with $\varepsilon=0$, $m=3$ and $p=5$. (a) and (b) show the experimental and theoretical Stokes parameters, respectively; and (c) the experimental (top) and theoretical (bottom) results for the reconstructed vector mode.
    %The resulting {\bf HLGVM} has $l=3$ and $q=1$.}
 %\label{fig:stokes}
%\end{figure}

\begin{figure}[h]
	\centering
	\includegraphics[width=.6\textwidth]{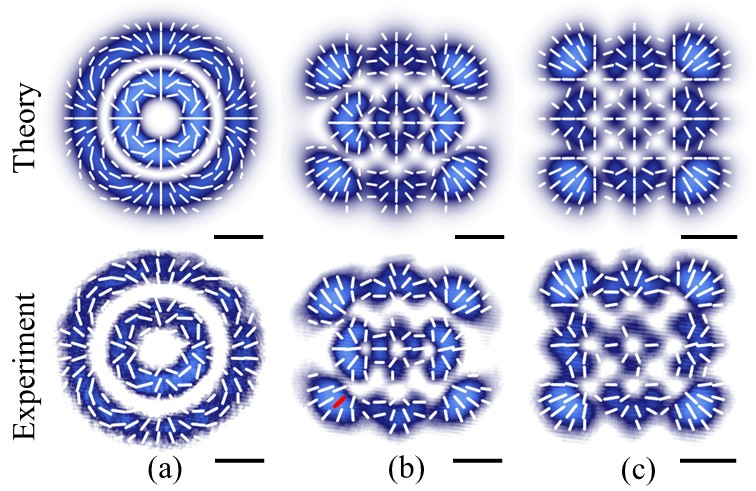}
	\caption{\small Numerical simulations (top) and experimental results (bottom) of the transverse polarisation distribution overlapped with the intensity profile of hIG vector modes with parameters $p=5$, $m=3$, $2\theta=\pi/2$, $2\delta=0$ and ellipticity (a) $\varepsilon=0$, (b) $\varepsilon=6$ and $\varepsilon=100$. Again, the scale bar on the bottom of each figure is 1 mm and.}
 \label{fig:vbeams}
\end{figure}

Fig.~\ref{fig:vbeams} shows the reconstructed polarisation distribution overlapped with the intensity profile of representative examples of hIG vector modes of parameters $p=5$, $m=3$, $2\theta=\pi/2$ and $2\delta=0$, numerical simulations on top and experimental results on the bottom. Fig.~\ref{fig:vbeams}(a) shows the case $\varepsilon=0$, Fig.~\ref{fig:vbeams}(b) shows the case $\varepsilon=6$ and Fig.~\ref{fig:vbeams}(c) the case $\varepsilon=100$. Notice the good agreement between the experiment and the numerical simulation.

\begin{figure*}[h]
	\centering
	\includegraphics[width=.8\textwidth]{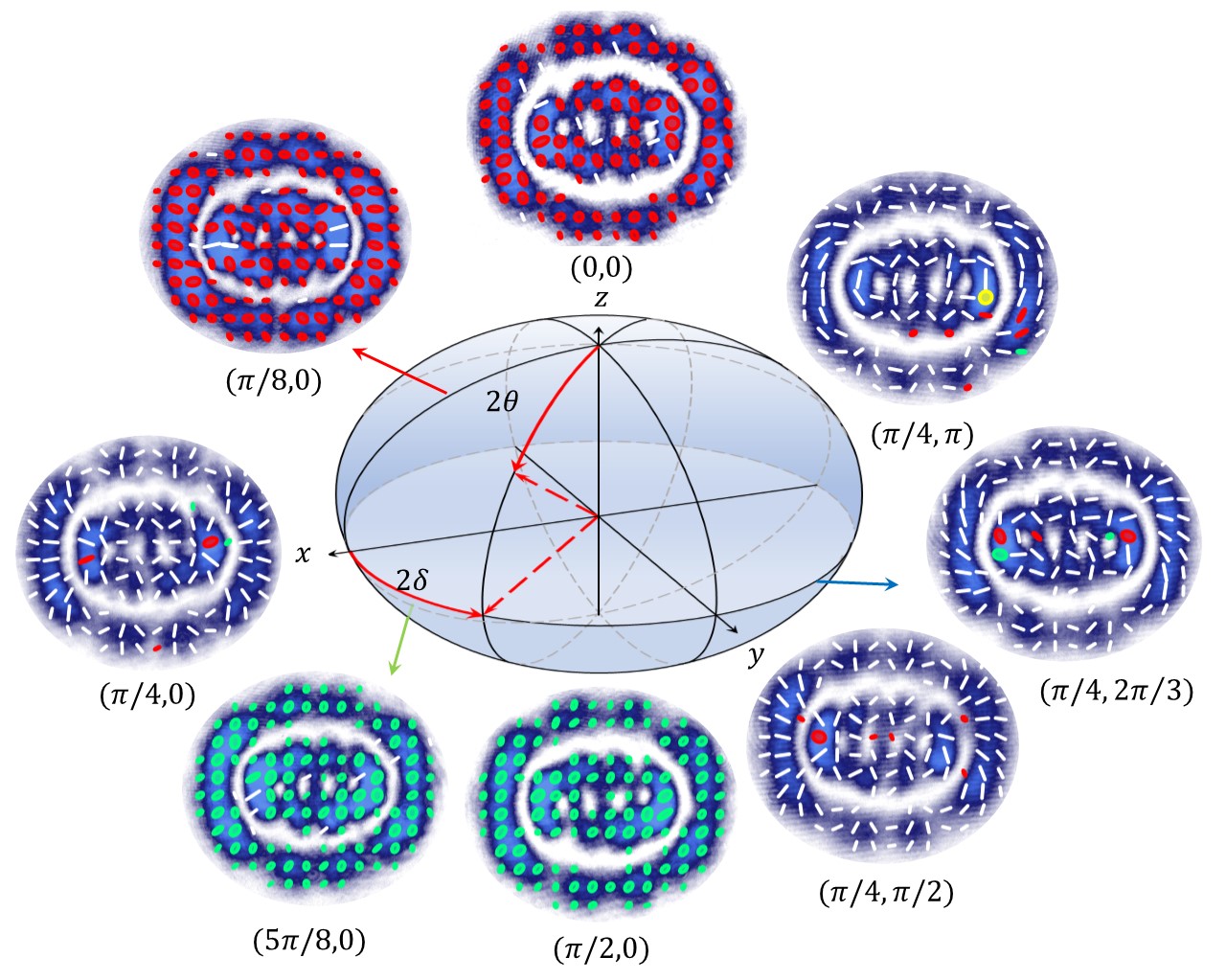}
	\caption{\small HOPE for experimental results of the {\bf hIG}$_{6,4,3}$ vector mode , with their corresponding coordinates $(2\delta,2\theta)$ shown below each one. Red ellipses refer to light with right- and green ones with left-handedness polarisation states; white segments represent linearly polarised light.}
 \label{fig:spheexp}
\end{figure*}

Lastly, in Fig.~\ref{fig:spheexp} we show a geometrical representation of experimentally generated hIG vector modes of parameters $p=6$, $m=4$ and elliptitity $\varepsilon=3$, on the surface of a HOPE of eccentricity ${\bf e}=0.39$. 
%This shows that under this new representation we can directly link the geometry of the modes with the geometry of the surface in which they are represented, breaking the paradigm of using spheres to represent vector modes and open the way to think in different alternatives like this HOPE.
%It preserves the advantages of the HOPS and permits the display of variations in the ellipticity of vector modes on the spheroid. Thus, in geometric terms, there exists a transitive representation that depends on $\varepsilon$.

%%%%%%%%%%%%%%%%%%%%%%%%%%%%%%%%%%%%%%%%%%%%%%%%%%
\section{Conclusions}
%%%%%%%%%%%%%%%%%%%%%%%%%%%%%%%%%%%%%%%%%%%%%%%%%%
Complex vector light fields, classically entangled in their spatial and polarisation degrees of freedom are reshaping the landscape of modern optics at both the fundamental and applied aspects, fuelling the development of numerous techniques for their generation and characterisation. Similarly, vector beams are pioneering diverse applications, which take full advantage of their unique properties. It is then crucial to develop tools for a more intuitive understanding of their properties. In the case of vector modes, this implies the understanding of the coupling between both degrees of freedom, polarisation, and spatial shape. The Poincar\'e sphere, first introduced by Poincar\'e in 1892, is one of the first examples that allowed a graphical representation in three dimensions of all polarisation states of light, as unique points on the surface of a unit-radius sphere. Similar representations have been proposed since then with a similar purpose, providing ways to visualize other properties of light. For example, the Poincar\'e sphere for OAM allows the representation of light beams with OAM as the superposition of Hermite-Gauss beams \cite{padgett1999poincare,galvez2003geometric}. More recently, the Higher-Order Poincar\'e Sphere (HOPS) provided the means for representing vector beams, non-separable in their spatial and polarisation degrees of freedom, also as points on the surface of a unit-radius sphere. Such representation allows visualising the evolution of scalar beams, which are located in the poles, to vector beams, which occupy the equator, as a function of a weighting factor. This representation adapts very well to vector beams with circular symmetry, such as Laguerre-Gauss or Bessel beams, but does not allow to visualise the shape evolution of light beams with, for example, elliptical shape, as is the case of hIG and Mathieu beams. Hence, in this manuscript, we put forward a generalisation of the HOPS, which we term Higher Order Poincar\'e Ellipsoid (HOPE). For simplicity, we choose an oblate spheroid that flattens in one direction. We provided a mathematical expression that relates the eccentricity of the oblate spheroid to the ellipticity of the hIG mode, establishing a one-to-one correspondence between hIG mode of ellipticity $\varepsilon$ and an oblate spheroid of eccentricity ${\bf e}$. The proposed relation, captures the asymptotic evolution in the shape of hIG modes, from Laguerre- (with circular shape) to Hermite-Gauss (with square shape), $\varepsilon=0$ and $\varepsilon\to\infty$, respectively. More precisely, we proposed an asymptotic dependence of  ${\bf e}$ in terms of $\varepsilon$ using the arctan function. In this way, the ellipticity interval of the hIG modes, $\varepsilon\in[0,\infty)$, is mapped onto the eccentricity interval of the ellipsoids, ${\bf e} \in [0,1/2)$. Hence, the proposed oblate spheroid, which we have termed Higher Order Poincar\'e Ellipsoid, allows the representation of hIG vector beams with unique ellipticities onto the surface of ellipsoids with unique eccentricities. Notice that, the HOPE becomes the HOPS for $\varepsilon=0$ and that the Stokes formalism is preserved, as expected. We anticipate that the HOPE will become a useful tool for the structured light community and will pave the path to more complete representations of the properties of light using geometric objects different from the traditional sphere.

\section*{Declarations}

{\bf Author contributions:} V.R.F, B.P.G and C.R.G. conceived the idea. D.D.S, E.M.S, V.R.G., B.P.G., and C.R.G. prepared the manuscript. C.R.G. Supervised the project.

\noindent
{\bf Funding:} No funding was received for conducting this study.

\noindent
{\bf Acknowledgements:} D.D.S. (CVU: 1159764) and E.M.S. (CVU: 742790) would like to acknowledge funding from Consejo Nacional de Humanidades Ciencia y Tecnologia (CONAHCYT) granted through their PhD's scholarships.

\noindent
{\bf Data availability:} The data presented in this study are available on request from the corresponding author.

\noindent
{\bf Consent to publish declaration:} Not applicable.

\noindent
{\bf Ethics approval and consent to participate:} Not applicable.

\noindent
{\bf Ethics and Consent to Participate declarations:} not applicable.

\noindent
{\bf Conflict of interest/Competing interests:} The authors have no conflicts of interest to declare that are relevant to the content of this article.

{\bf Submitted to the special issue "Structured lights: Properties and Applications" of Discover Applied Sciences.} 

%\bibliography{References}% common bib file
%% BioMed_Central_Bib_Style_v1.01

%% if required, the content of .bbl file can be included here once bbl is generated
%%\input sn-article.bbl

\end{document}